\documentclass[%
reprint,
nofootinbib,
nobibnotes,
amsmath,amssymb,
aps,
]{revtex4-2}

\usepackage{graphicx}
\usepackage{dcolumn}
\usepackage{bm}

\usepackage{amsmath,amssymb,amsbsy,amstext,amsthm,simplewick,amsfonts}
\usepackage{mathrsfs}
\usepackage{graphicx}
\usepackage{wrapfig}
\usepackage{upgreek}
\usepackage{bm} 
\usepackage{framed}
\usepackage{bbm}
\usepackage{textcomp}
\usepackage{adjustbox}
\usepackage{makecell}
\usepackage{tcolorbox}
\usepackage{empheq}
\usepackage[normalem]{ulem}
\usepackage{enumitem}
\usepackage{braket}
\usepackage{array}
\usepackage{dsfont}
\usepackage{physics}
\usepackage{ulem}
\usepackage{xcolor}
\usepackage{tocloft}
\usepackage{tikz}
\usepackage{ulem}
\usetikzlibrary{decorations.pathmorphing}

\usepackage{orcidlink}

\usepackage{hyperref}
\hypersetup{colorlinks=true,linkcolor=bluecyan,citecolor=red3,urlcolor=green3,pdfencoding=auto,linktocpage
}
\makeatletter
\let\pre@bibdata\@empty
\let\@bibdataout@init\relax
\makeatother

\definecolor{lightgreen}{cmyk}{0.2, 0, 0.2, 0.2}
\definecolor{lightgray}{cmyk}{0.1,0.2,0,0.1}
\definecolor{lightgray2}{cmyk}{0.1,0.1,0,0.1}
\definecolor{greyish2}{rgb}{.96,.96,.96}
\definecolor{bluecyan}{RGB}{0, 100, 200}
\definecolor{blue3}{RGB}{31,119,180}
\definecolor{red3}{RGB}{214,39,40}
\definecolor{orange3}{RGB}{255,127,14}
\definecolor{green3}{RGB}{44,160,44}
\definecolor{red2}{RGB}{255,0,0}
\definecolor{green2}{RGB}{0,170,0}
\definecolor{blue2}{RGB}{0,128,255}
\definecolor{magenta2}{RGB}{191,64,191}
\definecolor{purple2}{RGB}{112,48,160}
\definecolor{orange2}{RGB}{255,192,0}
\definecolor{red4}{RGB}{186,60,71}

\newcommand{\authoraffilsep}{\vspace{0.6em}}

\def\bfk{\mathbf{k}}

\def \sa{\mathsf{a}}
\def \sb{\mathsf{b}}

\def \dd{\mathrm{d}}
\def \pv{\theta}

\newcommand{\pFqcomma}{\mathchar"613B\mskip\pFqmuskip}

\newcommand*\pregFq[6][8]{%
	\begingroup
	\pFqmuskip=#1mu\relax
	\mathcode`\,=\string"8000
	\begingroup\lccode`\~=`\,
	\lowercase{\endgroup\let~}\pFqcomma
	{}_{#2}\mathbf{F}_{#3}{\left[\genfrac..{0pt}{}{#4}{#5};#6\right]}%
	\endgroup
}

\newmuskip\pFqmuskip
\newcommand*\pFq[6][8]{%
	\begingroup
	\pFqmuskip=#1mu\relax
	\mathcode`\,=\string"8000
	\begingroup\lccode`\~=`\,
	\lowercase{\endgroup\let~}\pFqcomma
	{}_{#2}F_{#3}{\left[\genfrac..{0pt}{}{#4}{#5};#6\right]}%
	\endgroup
}

\begin{document}
	
	\title{Exact Solutions for Chiral Gravitational Waves from Spin-2 Mixing}
	\author{Mohammad Ali Gorji\,\orcidlink{0000-0001-8848-2283}}
	\email{gorji@ibs.re.kr}
	\author{Yuhang Zhu\,\orcidlink{0000-0002-9771-3642}}
	\email{yhzhu@ibs.re.kr}
	
	\affiliation{\authoraffilsep
		Cosmology, Gravity, and Astroparticle Physics Group, Center for Theoretical Physics of the Universe, Institute for Basic Science (IBS), Daejeon, 34126, Korea\\
	}

	\date{\today}
	
\begin{abstract}
A spectator spin-2 field can mix linearly with the metric tensor perturbations. We study the coupled tensor system in an inflationary background, including a parity violating effect parametrised by $\theta$. Without expanding in the linear mixing between the two fields, we solve the system exactly for arbitrary mass while treating the mixing strength nonperturbatively. The mode functions of both helicities are constructed by acting with an operator-valued Gauss hypergeometric function ${}_2F_1$ on Whittaker modes, and the late-time power spectrum of each graviton helicity is expressed in closed form in terms of the generalised hypergeometric function ${}_3F_2$. We find that one helicity is exponentially enhanced, showing that strong mixing with a spectator spin-2 field can generate a large and highly chiral primordial gravitational wave spectrum. The degree of circular polarisation is bounded by $\tanh(\pi\theta)$, independently of the spin-2 mass and the mixing strength. The same solution also gives the late-time power spectrum of the spectator field and its cross spectrum with the graviton. In the weak mixing limit, our exact result reproduces the perturbative Schwinger-Keldysh result. These solutions provide a nonperturbative framework for studying primordial gravitational waves sourced by additional tensor modes.
\end{abstract}

\maketitle

\section{INTRODUCTION}\label{sec:intro}

Inflation provides the leading framework for describing the physics of the very early Universe \cite{Mukhanov:2005sc,Weinberg:2008zzc}. Its simplest realisations predict nearly scale-invariant, Gaussian, and adiabatic primordial fluctuations, in excellent agreement with observations of the cosmic microwave background (CMB) \cite{COBE:1992syq,WMAP:2012nax,Planck:2018jri}. Nevertheless, the particle content of the inflationary epoch remains largely unknown. Additional fields present during inflation can leave characteristic signatures in primordial correlation functions, whose momentum dependence may encode their masses and spins. This is the central idea of the cosmological collider \cite{Chen:2009zp,Baumann:2011nk,Noumi:2012vr,Arkani-Hamed:2015bza,Lee:2016vti}.

Additional spin-2 degrees of freedom are particularly interesting in this context. Massive spin-2 modes arise in massive gravity \cite{deRham:2010kj,Hinterbichler:2011tt,deRham:2014zqa}, bimetric gravity \cite{Hassan:2011zd,Hinterbichler:2012cn}, and metric-affine theories, where torsion and nonmetricity may contain helicity-2 excitations \cite{Aoki:2019snr,Aoki:2023sum}. Effective spin-2 modes can also appear in matter sectors. A well-studied example is provided by the tensor perturbations of an $SU(2)$ gauge field with an isotropic vacuum expectation value \cite{Maleknejad:2011jw,Maleknejad:2011sq}. Such models have been investigated extensively as sources of chiral gravitational waves (GWs) \cite{Adshead:2012kp,Maleknejad:2012fw,Adshead:2013nka,Namba:2013kia,Maleknejad:2014wsa,Obata:2016tmo,Maleknejad:2016qjz,Agrawal:2017awz,Agrawal:2018mrg,Domcke:2018rvv,Watanabe:2020ctz,Ishiwata:2021yne,Dimastrogiovanni:2023oid,Dimastrogiovanni:2025snj,Aoki:2025uwz}. This symmetry-breaking pattern is not restricted to gauge theories. It can also be realised by an isotropic triplet of vector fields with a global $SO(3)$ or $O(3)$ symmetry, which likewise contains effective spin-2 modes \cite{Emami:2016ldl,Firouzjahi:2018wlp,Gorji:2020vnh}. The coupling of these modes to the graviton can, for example, generate oscillatory features in the GW spectrum \cite{Gomez:2025jys}.

Even when its energy density is subdominant, a spectator sector can efficiently source primordial GWs through the linear mixing of its tensor modes with the metric tensor perturbations \cite{Gorji:2023ziy,Gorji:2023sil,Garriga:2025uko}. A model-independent description is provided by the effective field theory (EFT) of inflation \cite{Cheung:2007st}, extended to include fields with spin \cite{Bordin:2018pca} (see also \cite{Stefanyszyn:2023qov,Garriga:2025uko,Cheung:2025dmc}). At the lowest derivative order, the mixing is generated by the operator $\sigma^{ij}\delta K_{ij}$, where $\sigma_{ij}$ is the additional spin-2 field and $\delta K_{ij}$ is the perturbation of the extrinsic curvature of constant-time hypersurfaces. Since $\sigma_{ij}$ transforms only under the unbroken spatial rotations, rather than the full de Sitter isometry group, the usual Higuchi bound can be evaded \cite{Higuchi:1986py,Bordin:2016ruc}. The spin-2 field can therefore remain light and survive on superhorizon scales \cite{Bordin:2018pca}.

The same EFT admits parity-odd operators in the massive spin-2 sector \cite{Stefanyszyn:2023qov,Garriga:2025uko}. These operators distinguish the two helicities of the additional tensor field, and the resulting asymmetry is transferred to the graviton through the linear mixing, generating a chiral primordial GW spectrum. Related mechanisms producing helicity asymmetries and potentially large signals from massive spin-2 and higher spin fields have also been studied in \cite{Tong:2022cdz,An:2025mdb}.  Such chirality may be probed through the parity-odd $TB$ and $EB$ correlations of the CMB \cite{Thorne:2017jft,LiteBIRD:2023zmo}, the intrinsic alignments of galaxies \cite{Okumura:2024xnd,Mikura:2025vaj}, and GW observations \cite{Caprini:2018mtu,LISA:2017pwj,Orlando:2020oko}. Chiral GWs produced before the electroweak epoch can also be constrained by the observed baryon asymmetry through the gravitational chiral anomaly \cite{Gorji:2026tln}.

The quadratic mixing between the two spin-2 fields is usually treated perturbatively as an interaction vertex, with correlation functions computed order by order using the in-in or Schwinger-Keldysh (SK) formalism \cite{Chen:2017ryl,Firouzjahi:2018wlp,Gorji:2020vnh,Qin:2023closed}. This treatment applies only when the mixing is sufficiently weak, while beyond the perturbative regime the coupled linear system must be solved nonperturbatively. The problem is nontrivial because the mass and mixing terms cannot be diagonalised simultaneously, and no constant field rotation decouples the system throughout its evolution. Recently, exact solutions for arbitrary mixing strength and isocurvature mass were obtained for coupled scalar perturbations in the quasi-de Sitter background \cite{Huenupi:2026abj}. 

In this work, we extend the exact nonperturbative treatment of coupled perturbations to the tensor sector, including a parity violating effect parametrised by $\theta$. Starting from a special value of the spin-2 mass for which the eigenchannels decouple and the solutions reduce to Whittaker functions, we construct the mode functions for arbitrary mass and mixing strength using an operator-valued Gauss hypergeometric function. This construction allows us to derive the \textit{exact} late-time power spectrum for each graviton helicity in closed form. We find that strong mixing produces a large and highly chiral primordial GW spectrum, with the degree of circular polarisation bounded by $\tanh(\pi\theta)$, and verify the exact result through direct numerical evolution and its weak mixing and parity symmetric limits. The same solution also gives the spectator power spectrum and its cross spectrum with the graviton at late times, allowing us to examine the helicity asymmetry of the field that sources the chiral GW signal.

The paper is organised as follows. In Sec.~\ref{sec:setup}, we introduce the coupled spin-2 system and its quantisation. Sec.~\ref{sec:special} presents the exact solution at the special mass $\Delta_t=0$, for which the two eigenchannels decouple, and Sec.~\ref{sec:general} extends the construction to an arbitrary mass. In Sec.~\ref{sec:spectrum}, we derive the late-time power spectrum of two graviton helicities in closed form, together with the leading late-time spectator power spectrum and its cross spectrum with the graviton. Sec.~\ref{sec:limits} examines the special-mass, parity-symmetric, and weak-mixing limits, including the bound on the degree of circular polarisation and the single-exchange SK result. We conclude in Sec.~\ref{sec:summary}. The derivation of the late-time spectator mode is presented in App.~\ref{app:sigma-cross-proof}.

\section{SET-UP}\label{sec:setup}
We begin by specifying the coupled spin-2 system that will be the focus of this work. One of the two spin-2 fields is the usual tensor perturbation of the spatial metric, denoted by $h_{ij}$. In terms of the canonically normalised field $\gamma_{ij}=M_{\text{pl}} h_{ij}/\sqrt{2}$, its free action takes the standard form
\begin{align}
S^{(2)}_{\gamma}=\frac{1}{4}\int \dd\tau \dd^3x\,a^2(\tau) \Big[\gamma'_{ij}\gamma'_{ij}-\partial_{l}\gamma_{ij}\partial_{l}\gamma_{ij}\Big]\,,
\end{align}
where we work on an inflationary background with scale factor $a(\tau)=-1/(H\tau)$. Here $H$ denotes the Hubble scale and $\tau$ is conformal time, while spatial indices are contracted with the Kronecker delta $\delta_{ij}$. Throughout this work, unless otherwise specified, a prime denotes differentiation with respect to the argument of the function. Imposing the transverse and traceless conditions on the symmetric tensor $\gamma_{ij}$ leaves only two degrees of freedom.
	
The second field is a massive spectator spin-2 field $\sigma_{ij}$, whose quadratic action we take to be \cite{Bordin:2018pca,Stefanyszyn:2023qov,Garriga:2025uko,Cheung:2025dmc}
\begin{align}
	S^{(2)}_{\sigma}=&\frac{1}{4}\int \dd\tau \dd^3x\,a^2(\tau) \Big[\sigma'_{ij}\sigma'_{ij}-\partial_{l}\sigma_{ij}\partial_{l}\sigma_{ij}\nonumber\\
	&-\frac{m^2-\alpha^2}{\tau^2}\sigma_{ij}\sigma_{ij}+2\frac{\pv}{\tau} \epsilon_{i k m}\sigma_{ij}\partial_k \sigma_{mj} \Big]\,. \label{eq: sigma_action}
\end{align}
To simplify the expressions below, we parametrise the coefficient of the mass term as $m^2-\alpha^2$, where $m$ and $\alpha$ are expressed in Hubble units and $\alpha$ will shortly be identified with the strength of the quadratic mixing. The degrees of freedom considered here are classified under spatial rotations and carry only spatial indices, following the construction of \cite{Bordin:2018pca} (see also \cite{Bordin:2016ruc,Stefanyszyn:2023qov,Garriga:2025uko,Cheung:2025dmc}). Their coupling to the preferred inflationary foliation breaks the full de Sitter isometries, so the spin-2 field is not subject to the usual Higuchi bound \cite{Higuchi:1986py} and can therefore remain light \cite{Bordin:2016ruc}. We also include a parity-odd contribution through the final term in \eqref{eq: sigma_action}, with its strength controlled by the dimensionless parameter $\pv$, causing the two tensor helicities to evolve differently and thereby sourcing chiral primordial GWs.

Although $\sigma_{ij}$ is symmetric and traceless, it is \textit{not transverse} and therefore contains five helicity components. At quadratic order, the metric tensor perturbations couple only to the helicity-2 component of $\sigma_{ij}$, and in this sector the EFT fixes the mixing between the two spin-2 fields uniquely at leading order in the derivative expansion, through the operator $\sigma^{ij}\delta{K}_{ij}$ \cite{Bordin:2018pca},
\begin{align}
	S^{(2)}_{\text{int}}=-\frac{1}{2}\int \dd\tau  \dd^3 x\,a^2(\tau)\frac{\alpha}{\tau}\,\gamma'_{ij} \sigma_{ij}\,, \label{eq:int}
\end{align}
one may ask whether, at the same order in the derivative expansion, the time derivative can instead be replaced by a spatial derivative contracted with the Levi-Civita tensor, giving a parity-odd interaction such as $\epsilon_{ikm}\sigma_{ij}\partial_k\gamma_{mj}$ and hence an additional contribution to chiral GW production. However, the form of the allowed interactions is fixed by the pattern of symmetry breaking. For metric perturbations, the relevant EFT building block at this derivative order is $\delta K_{ij}\propto\gamma'_{ij}$, whereas a bare spatial derivative of $\gamma_{ij}$ is not allowed.\footnote{This differs from models with $SU(2)$ gauge fields, which preserve a diagonal combination of internal and spatial rotations \cite{Maleknejad:2011sq,Aoki:2025uwz}. In that case, parity-odd operators that directly couple the graviton to the additional tensor mode can appear already at leading derivative order \cite{Agrawal:2017awz,Agrawal:2018mrg,Aoki:2025uwz}. The two frameworks therefore lead to structurally different chiral interactions.}

The dimensionless parameter $\alpha$ in \eqref{eq:int} controls the strength of this quadratic mixing, and our aim is to solve the coupled system \textit{nonperturbatively} in $\alpha$ rather than treat the mixing as a perturbative interaction. The same operator $\sigma^{ij}\delta{K}_{ij}$ also mixes the helicity-0 component of $\sigma_{ij}$ with the scalar perturbation \cite{Bordin:2018pca}. Constraints from this sector depend on the helicity-0 dynamics and its propagation speed, neither of which directly enters the tensor system studied here. We therefore treat $\alpha$ as a free parameter and derive the exact solution of the tensor system for arbitrary mixing strength, leaving the allowed range of $\alpha$ to be determined within a specific model. In this sense, our system provides the tensor counterpart of the scalar model studied in \cite{Huenupi:2026abj} and extends it nontrivially by including parity violation. The resulting solution can serve as a theoretical framework for more detailed phenomenological studies of the mixing between gravitons and spin-2 fields in future work.

We now decompose the fields into helicity modes and transform to momentum space as
\begin{align}
	\gamma_{ij}(\tau,\bf{x})
	&=\sum_{\lambda=\pm2}\int_{\bf{k}}\,e^{i\bf{k}\cdot\bf{x}}e_{ij}^{(\lambda)}(\hat{\bf{k}})
	\gamma_\lambda(\tau,\bf{k}) \,,
	\\
	\sigma_{ij}^{\mathrm{TT}}(\tau,\bf{x})&=\sum_{\lambda=\pm2}
	\int_{\bf{k}}\,e^{i\bf{k}\cdot\bf{x}}e_{ij}^{(\lambda)}(\hat{\bf{k}})\sigma_\lambda(\tau,\bf{k}) \,,
\end{align}
where $\int_{\bfk}\equiv \int \frac{\dd^3k}{(2\pi)^3}$ and the superscript $\mathrm{TT}$ denotes the transverse and traceless part of $\sigma_{ij}$, which is the only part entering \eqref{eq:int}.  We will therefore omit this superscript in what follows, with the understanding that $\sigma_\lambda$ always refers to the helicities $\lambda=\pm2$.
	
For the polarisation tensors, we choose the reality condition and normalisation as
\begin{align}
	\left[e_{ij}^{(\lambda)}(\hat{\bf{k}})\right]^*
	=e_{ij}^{(\lambda)}(-\hat{\bf{k}}) \,,~~
	e_{ij}^{(\lambda)}(\hat{\bf{k}})e_{ij}^{(\lambda')}(-\hat{\bf{k}})
	=2\delta_{\lambda\lambda'} \,,
\end{align}
together with the helicity identity
\begin{align}
	i\epsilon_{i l m}k_le_{mj}^{(\lambda)}(\hat{\bf{k}})&=\frac{\lambda}{2}k\,e_{ij}^{(\lambda)}(\hat{\bf{k}}) \,.
\end{align}
Using the shorthand $A_\lambda B_\lambda\equiv A_\lambda(\tau,{\bf k})B_\lambda(\tau,-{\bf k})$, the full quadratic action of the helicity-2 sector in momentum space reads
\begin{align}
\begin{split}
	S^{(2)}&=\frac{1}{2}\sum_{\lambda=\pm2}\int_{\bfk}\dd\tau\,a^2(\tau)\Big[\gamma_\lambda'^2-k^2\gamma_\lambda^2-\frac{2\alpha}{\tau}\gamma'_{\lambda}\sigma_{\lambda}\\
	&~~~~~~~~+\sigma_\lambda'^2-\left(k^2-\frac{\lambda \pv k}{\tau}+\frac{m^2-\alpha^2}{\tau^2}\right)
	\sigma_\lambda^2\,\Big]\,.
\end{split}
\label{eq:full_action}
\end{align}
For $\pv=0$, the helicity dependence drops out and each helicity sector is mathematically equivalent to the coupled scalar system \cite{Huenupi:2026abj}, whereas $\pv\neq0$ leads to genuinely helicity-dependent dynamics.
    
To quantise the system, we first find the canonical momenta as $\Pi_{\gamma,\lambda}=a^2\left(\gamma_\lambda'-\frac{\alpha}{\tau}\sigma_\lambda\right)$ and $\Pi_{\sigma,\lambda}=a^2\sigma_\lambda'$. They obey the equal-time commutation relations
\begin{align}
	\Big[\gamma_\lambda(\tau,\bfk),\Pi_{\gamma,\lambda'}(\tau,\bf{k}')\Big]&=i\,(2\pi)^3\delta_{\lambda\lambda'}\delta^{(3)}(\bf{k}+\bf{k}') \,,
	\\
	\Big[\sigma_\lambda(\tau,\bfk),\Pi_{\sigma,\lambda'}(\tau,\bfk')\Big]
	&=i\,(2\pi)^3\delta_{\lambda\lambda'}
	\delta^{(3)}(\bfk+\bfk') \,.
\end{align}
Since the coupled system contains two fields, each helicity requires two independent pairs of creation and annihilation operators. We label the corresponding oscillator modes by $\sa=\pm1$ and expand the fields as
\begin{align}
	&\gamma_{\lambda}(\tau,\bfk)=\sum_{\sa=\pm1} \left[\gamma_{\lambda\sa}(\tau,k)\,a_{\lambda\sa}(\bfk)+\gamma^*_{\lambda\sa}(\tau,k)\, a^\dagger_{\lambda\sa} (-\bfk)\right]\,,\nonumber\\
	&\sigma_{\lambda}(\tau,\bfk)=\sum_{\sa=\pm1} \left[\sigma_{\lambda\sa}(\tau,k)\,a_{\lambda\sa}(\bfk)+\sigma^*_{\lambda\sa}(\tau,k)\, a^\dagger_{\lambda\sa} (-\bfk)\right]\,,
\end{align}
where
\begin{align}
	\left[a_{\lambda\sa}(\bfk),a_{\lambda'\sa'}^\dagger(\bfk')\right]
	&=(2\pi)^3\delta_{\lambda\lambda'}\delta_{\sa\sa'}
	\delta^{(3)}(\bfk-\bfk') \,.
\end{align}
		
\section{Solution at the Special Mass}\label{sec:special}
Varying the action \eqref{eq:full_action} with respect to the fields gives the equations of motion
\begin{align}
	\gamma_\lambda''-\frac{2}{z}\gamma_\lambda'+\gamma_\lambda-\frac{\alpha}{z}
	\left(\sigma_\lambda'-\frac{3}{z}\sigma_\lambda\right)&=0 \,,\\
	\sigma_\lambda''-\frac{2}{z}\sigma_\lambda'
	+\left(1+\frac{\lambda\pv}{z}+\frac{m^2-\alpha^2}{z^2}\right)\sigma_\lambda+\frac{\alpha}{z}\gamma_\lambda' &=0\,,
\end{align}
where we have defined the dimensionless time variable $z=-k\tau$. To make the underlying structure more transparent, we remove the friction terms using the redefinition
\begin{align}
	X_\lambda =-\frac{1}{z}\left(\gamma_\lambda'-\frac{\alpha}{z}\sigma_\lambda\right) \,,\qquad
	Y_\lambda =-\frac{\sigma_\lambda}{z} \,,
\end{align}
using the equations of motion, the original fields can be recovered as
\begin{align}
	\gamma_\lambda =zX_\lambda'-X_\lambda \,,
	\qquad
	\sigma_\lambda =-zY_\lambda \,. \label{eq: inverse_relation}
\end{align}
Substituting these relations back into the equations of motion gives
\begin{align}
	{\bm{\Psi}}_\lambda''+\left(\mathsf I+\frac{{\mathsf{K}}_\lambda}{z}
	+\frac{\Delta_t{\mathsf{M}}}{z^2}\right)\bm{\Psi}_\lambda =0 \,, \label{eq: EoM_Psi}
\end{align}
with ${\bm{\Psi}}_\lambda \equiv ( X_\lambda\,\,Y_\lambda)^{T}$ and 
\begin{align}
	\mathsf K_\lambda =\begin{pmatrix}0&\alpha\\\alpha&\lambda\pv\end{pmatrix} \,, \qquad
	\mathsf M =\begin{pmatrix}0&0\\0&1\end{pmatrix} \,.
\end{align}
The mass dependence is given by the parameter
\begin{align}
	\Delta_t \equiv m^2-2\equiv\frac{1}{4}-\nu_t^2 \,.
\end{align}
One may think that two oscillators can be decoupled by performing a rotation. However, a constant rotation cannot diagonalise the full system when $\Delta_t\neq0$ since the two matrices do not commute for $\alpha\neq0$
	\begin{align}
		\left[\mathsf K_\lambda,\mathsf M\right] &=\alpha
		\begin{pmatrix}
			0&1\\
			-1&0
		\end{pmatrix} \,.
	\end{align}
A time-dependent rotation does not provide a simplification either, since it introduces additional derivative couplings. 
	
To construct the exact solutions, we begin with the reference case $\Delta_t=0$, for which the $z^{-2}$ term vanishes and the whole problem simplifies dramatically. We can then easily diagonalise the complete system by finding the two eigenvalues of the matrix $\mathsf{K}_{\lambda}$, which we denote by $q_{\lambda\sa}$ with $\sa=\pm1$:
\begin{align}
	q_{\lambda\sa}=\frac{\lambda\pv}{2}+\sa\rho \,, \qquad
	\rho =\sqrt{\alpha^2+{\pv^2}} \,. \label{eq: def_q}
\end{align}
The corresponding unit eigenvectors may be written as
\begin{align}
	{\bm{v}}_{\lambda\sa}&=\sqrt{w_{\lambda\sa}}
	\begin{pmatrix}
	1\\
	{q_{\lambda\sa}}/{\alpha}
	\end{pmatrix} \,,\\
	w_{\lambda\sa}&=\frac{\alpha^2}{\alpha^2+q_{\lambda\sa}^2} =\frac{1}{2}\left(1-\frac{\sa\lambda\pv}{2\rho}\right) \,.
\end{align}
We choose the two oscillator modes introduced above to coincide with these eigenchannels. A mode in channel $\sa$ can therefore be written as
\begin{align}
	{\bm\Psi}_{\lambda\sa}(z)={\bm {v}}_{\lambda\sa}u_{q_{\lambda\sa}}(z) \,, \label{eq: eigendecoms}
\end{align}
put this back to the equation of motion with $\Delta_t=0$ gives
\begin{align}
	\left[u_{q_{\lambda\sa}}''+\left(1+\frac{q_{\lambda\sa}}{z}\right)u_{q_{\lambda\sa}}\right] \bm v_{\lambda\sa}=0\,.
\end{align}
Since $\bm v_{\lambda\sa}$ is a constant vector, the time-dependent bracket must vanish, leaving a single second-order equation for $u_{q_{\lambda\sa}}$. Its positive frequency solution is
\begin{align}
\begin{split}
	u_{q_{\lambda \sa}}(z)&=\mathcal{C}(k)\, e^{-\pi q_{\lambda\sa}/4}\,W_{iq_{\lambda\sa}/2,\,1/2}\left(-2iz\right) \,\\
	&=\mathcal{C}(k)\,e^{-\pi q_{\lambda \sa}/4} e^{iz}\,U\left(-{iq_{\lambda \sa}}/{2},0,-2iz\right)\,,
\end{split}
\label{eq: sol_of_u}
\end{align}
where the first line uses the Whittaker-$W$ function, while the second gives the equivalent form in terms of the confluent hypergeometric function $U$, allowing a direct comparison with the scalar results obtained in \cite{Huenupi:2026abj}.
	
As usual, the remaining factor $\mathcal{C}(k)$ is fixed by imposing the Bunch-Davies (BD) condition in the far past. Under $z\to\infty$, the Whittaker function becomes
\begin{align}
	\lim_{z\to\infty}e^{-\pi q/4}\,W_{iq/2,\,1/2}\left(-2iz\right)
	&\sim e^{iz}\left(2z\right)^{iq/2}\,.
\end{align}
Since $z=-k\tau$, the factor $e^{iz}=e^{-ik\tau}$ is the positive frequency mode. Here the additional factor $(2z)^{iq/2}$ is just a phase accumulated from the $q/z$ term and leaves the amplitude of the mode unchanged.
	
In the conventional BD basis, the two independent positive frequency modes usually are chosen in the far past to excite $\gamma_\lambda$ and $\sigma_\lambda$ separately. The eigenchannel modes used here are unitary linear combinations of this pair. Combining \eqref{eq: eigendecoms} with \eqref{eq: inverse_relation}, their behaviour in the far past is 
\begin{align}
	\lim_{z\to\infty}
		\begin{pmatrix}
			\gamma_{\lambda\sa}\\
			\sigma_{\lambda\sa}
		\end{pmatrix}
		&\sim \mathcal{C}\,z\,e^{iz}\left(2z\right)^{iq_{\lambda\sa}/2}
		\sqrt{w_{\lambda\sa}}
		\begin{pmatrix}
			i\\
			-q_{\lambda\sa}/\alpha
		\end{pmatrix} \,. \label{eq: initial_condition}
\end{align}
The two column vectors above, corresponding to $\sa=\pm1$, are orthonormal, since $q_{\lambda+}q_{\lambda-}=\det\mathsf K_\lambda=-\alpha^2$, so the eigenchannel modes and the conventional BD modes are related by a unitary transformation. The creation and annihilation operators transform inversely, leaving the field expansion and the canonical commutation relations unchanged. We may therefore impose the BD normalisation directly in the eigenchannel basis. Since each vector has unit norm, matching the common amplitude to $1/(a\sqrt{2k})$ fixes the coefficient up to an irrelevant overall phase as
\begin{align}
	\mathcal{C}(k)=\frac{H}{\sqrt{2k^3}} \,.
\end{align}
The power spectrum is obtained by summing over both independent oscillator modes and it is then invariant under the same unitary transformation. We have now obtained the complete solution to the coupled system in the special case $\Delta_t=0$.
	
\section{Solution of General Mass}\label{sec:general}
For a general mass, the two coupled second-order equations in \eqref{eq: EoM_Psi} can be combined into a single fourth-order equation. For example, we can solve the first equation for $Y_\lambda$, which gives $Y_\lambda=-\frac{z}{\alpha}\left(X_\lambda''+X_\lambda\right)$, and then put this relation into the second equation, yielding
\begin{align}
	\mathcal{L}_{\Delta_t}X_\lambda=0 \,,
\end{align}
where the fourth-order differential operator reads
\begin{align}
	\mathcal L_{\Delta_t}&=\left(\mathcal A+q_{\lambda+}\right)\left(\mathcal A+q_{\lambda-}\right)+\Delta_t\,\mathcal{G}(\hat{x}) \,.
\end{align}
We have introduced the shorthand notation
\begin{align}
	\mathcal A &\equiv z\left(\partial_z^2+1\right) \,,
	\qquad
	\mathcal{G}(\hat{x}) \equiv4{\hat{x}}\left(1-\hat{x}\right)=\partial_z^2+1 \,,
\end{align}
with
\begin{align}
	\hat{x}\equiv\frac{1+i\partial_{z}}{2} \,,
\end{align}
and $q_{\lambda\pm}$ given by \eqref{eq: def_q}. With this notation, the reference solution we found in \eqref{eq: sol_of_u} then satisfies 
\begin{align}
	\left(\mathcal A+q_{\lambda\sa}\right)u_{q_{\lambda\sa}}=0\,.
\end{align}
To extend the reference solutions to arbitrary $\Delta_t$, we act on each mode with an appropriate operator $\mathcal D_\sa$ and make the following ansatz for the general mass solutions 
\begin{align}
	X_{\lambda\sa} &=\sqrt{w_{\lambda\sa}}\,\mathcal{D}_\sa(\hat{x})\,u_{q_{\lambda\sa}}\,.\label{eq: general_sol}
\end{align}
For determining the explicit form of this operator, we first use $\left[z,\hat{x}\right]=-i/2$ to derive several commutation relations
\begin{align}
\begin{split}
	\left[z,\mathcal{D}_{\sa}(\hat{x})\right]&=-\frac{i}{2}\mathcal{D}'_{\sa}(\hat{x})\,,\\
	\left[\mathcal{A},\mathcal{D}_{\sa}(\hat{x})\right]&=-\frac{i}{2}\mathcal{G}(\hat{x})\,\mathcal{D}'_{\sa}(\hat{x})\,,
\end{split}
\end{align}
here $\mathcal{D}_\sa'(x)=\dd \mathcal{D}_\sa(x)/\dd x$ denotes the derivative with respect to the operator argument $\hat{x}$. Applying these commutation relations, we get
\begin{align}
\mathcal{L}_{\Delta_t}\mathcal{D}_\sa(\hat{x})u_{q_{\lambda\sa}}&=-\mathcal{G}(\hat{x})\Big[\hat{x}\left(1-\hat{x}\right)\mathcal{D}_\sa''\nonumber\\
	&+\left(1-i\sa\rho-2\hat{x}\right)\mathcal{D}_\sa'
	-\Delta_t\mathcal{D}_\sa
	\Big]u_{q_{\lambda\sa}} \,.
\end{align}
In order to solve the original equation we require the expression in brackets to vanish, 
\begin{align}
	\hat{x}\left(1-\hat{x}\right)\mathcal{D}_\sa''+\left(1-i\sa\rho-2\hat{x}\right)\mathcal{D}_\sa'-\Delta_t\mathcal{D}_\sa=0\,.
\end{align}
The solution satisfying $\mathcal{D}_\sa(0)=1$ is
\begin{align}\label{eq: D-def}
	\mathcal{D}_\sa(\hat{x})=\pFq{2}{1}{\frac{1}{2}-\nu_t,\frac{1}{2}+\nu_t}{1-i\sa\rho}{\hat{x}}\,.
\end{align}
When $\nu_t=1/2$, corresponding to the special mass case considered before, as expected, this operator reduces to the identity $\left.\mathcal{D}_\sa(\hat{x})\right|_{\nu_t=1/2}=1$. The fourth-order equation has four independent solutions for each helicity, and since we have already constructed the two modes $X_{\lambda\sa}$ with $\sa=\pm1$, the reality of $\mathcal{L}_{\Delta_t}$ implies that the remaining two are just their complex conjugates, so that $\{X_{\lambda+},X_{\lambda-},X^*_{\lambda+},X^*_{\lambda-}\}$ forms a complete basis of solutions.	
	
Although \eqref{eq: general_sol} and \eqref{eq: D-def} solve the equations of motion for an arbitrary mass, it remains to verify that they carry the correct BD normalisation. This can be directly checked using the normalisation coefficient already fixed in the $\Delta_t=0$ case. To see this, we note that the mass term in \eqref{eq: EoM_Psi} becomes negligible in the far past, so the modes have the same asymptotic behaviour as in the $\Delta_t=0$ case, for which the amplitude has already been correctly fixed. Since $\hat{x}\,u_{q_{\lambda\sa}}=\mathcal{O}(z^{-1})$, then $\mathcal D_\sa(\hat{x})u_{q_{\lambda\sa}}=u_{q_{\lambda\sa}}$ in the limit $z\to \infty$, and \eqref{eq: general_sol} therefore gives the correctly normalised positive frequency solutions for a general mass. The other component can be obtained directly from the relation $\alpha Y_\lambda=-\mathcal A X_\lambda$.  This completes the construction of the \textit{exact} solution of the coupled system \eqref{eq: EoM_Psi}.
	
\section{late-time limit and power spectrum}\label{sec:spectrum}
\subsection{Graviton Power Spectrum}
\begin{figure}[htp]
	\centering
	\hspace{-1cm}
	\includegraphics[width=0.45\textwidth]{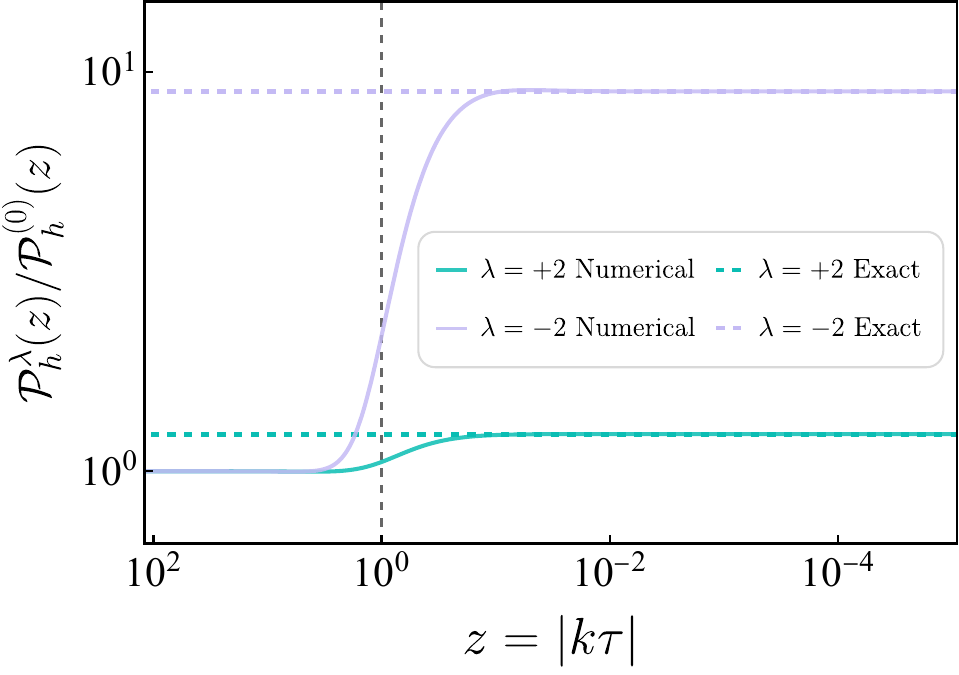}
	\caption{Time evolution of the dimensionless tensor power spectrum for the two helicities $\lambda=\pm 2$. The spectrum is normalised by the standard free graviton spectrum per helicity $\mathcal{P}^{(0)}_h(z)=H^2(1+z^2)/\pi^2M_{\text{pl}}^2$. The solid curves are obtained by numerically solving the coupled equations, while the dashed lines follow from the exact late-time expression. We take $\alpha=1$, $m=2$ and $\pv=4/3$.\label{fig: tensorPS}}
	\end{figure}
With the exact solution in hand, we now turn to the tensor power spectrum generated by this nonperturbative mixing. Since the mode functions are constructed by applying an operator to the reference solutions, their superhorizon behaviour is most readily obtained from the integral representation of the confluent hypergeometric function $U$, which gives \cite{NIST:DLMF}
\begin{align}
\begin{split}
	u_{q_{\lambda\sa}}(z)&=\frac{H}{\sqrt{2k^3}} \frac{e^{-\pi q_{\lambda\sa}/4}e^{iz}}
	{\Gamma\left(-iq_{\lambda\sa}/2\right)}\\
	&\times\int_0^\infty\dd t\,
	t^{-1-iq_{\lambda\sa}/2}\left(1+t\right)^{-1+iq_{\lambda\sa}/2}e^{2izt} \,.
\end{split}
\label{eq:u-int-rep}
\end{align}
Strictly speaking, this integral should be understood by analytic continuation from $\text{Im}(q_{\lambda\sa})>0$, with the usual $i\epsilon$ prescription regulating the oscillatory behaviour at large $t$. In this representation, all the $z$-dependence is carried by the exponential, and since $\hat{x}\,e^{iz(1+2t)}=-t\,e^{iz(1+2t)}$, the action of the operator can be evaluated simply by replacing $\hat{x}$ with $-t$ throughout, yielding
\begin{align}
\begin{split}
	&\lim_{z\to0}\mathcal D_{\sa}(\hat{x})u_{q_{\lambda\sa}} (z)=\frac{H}{\sqrt{2k^3}} \frac{e^{-\pi q_{\lambda\sa}/4}}{\Gamma\left(-iq_{\lambda\sa}/2\right)}
	\\
	&\times\int_0^\infty\frac{\dd t}{t^2}\,\pFq{2}{1}{\frac{1}{2}-\nu_t,\frac{1}{2}+\nu_t}{1-i\sa\rho}{-t}\,
	\left(\frac{t}{1+t}\right)^{1-i{q_{\lambda\sa}}/2}\,,
\end{split}
\end{align}
where the superhorizon limit has been taken by sending the exponential to unity. The remaining integral can then be evaluated exactly, giving the closed form expression as
\begin{align}
	\lim_{z\to0}X_{\lambda\sa}(z)
	&=\frac{H}{\sqrt{2k^3}}\sqrt{w_{\lambda\sa}}\,\mathcal T_{\lambda\sa} \,,\label{eq:X-solution}
\end{align}
with the dimensionless coefficient $\mathcal T_{\lambda\sa}$ is given by
\begin{align}
	\mathcal T_{\lambda\sa} &= e^{-\pi q_{\lambda\sa}/4}
	\frac{\Gamma\left(\frac{3}{2}-\nu_t\right)}{\Gamma\left(\frac{3}{2}-\nu_t-\frac{iq_{\lambda\sa}}{2}\right)} \nonumber
	\\
	&~~\times\pFq{3}{2}{\frac{1}{2}-\nu_t,\frac{1}{2}-\nu_t-i\sa\rho,-\frac{iq_{\lambda\sa}}{2}}{1-i\sa\rho,\frac{3}{2}-\nu_t-\frac{iq_{\lambda\sa}}{2}} 
	{1} \,. 
\label{eq: 3F2expression}
\end{align}
This expression is the central result of this work. The coefficient $\mathcal T_{\lambda\sa}$ determines the late-time limit of $X_{\lambda\sa}$ through \eqref{eq: general_sol}. The corresponding canonically normalised graviton mode $\gamma_{\lambda\sa}$ then follows from \eqref{eq: inverse_relation}, giving $\gamma_{\lambda\sa}(0)=-(H/\sqrt{2k^3})\sqrt{w_{\lambda\sa}}\mathcal T_{\lambda\sa}$. Finally, the dimensionless power spectrum for $h_\lambda=\sqrt{2}\,\gamma_\lambda/M_{\mathrm{pl}}$, for each helicity, is defined by
\begin{align}
	\langle h_\lambda({\bf{k}})h_{\lambda'}({\bf{k}'})\rangle
	&=(2\pi)^3\delta_{\lambda\lambda'}
	\delta^{(3)}({\bf{k}}+{\bf{k}'})\frac{\pi^2}{k^3}
	\mathcal P_h^\lambda(k)\,. \label{eq: h-powerS}
\end{align}
Denoting the late-time free graviton power spectrum per helicity by
\begin{align}\label{eq:PS-free}
\mathcal P_h^{(0)}\equiv\mathcal P_h^{(0)}(0)=\frac{H^2}{\pi^2M_{\mathrm{pl}}^2} \,,
\end{align}
the mode expansion above gives
\begin{align}
	\mathcal P_h^\lambda &=\frac{2k^3}{\pi^2M_{\mathrm{pl}}^2}
	\sum_{\sa=\pm}
	\left|\gamma_{\lambda\sa}\right|^2
	={\mathcal P}_h^{(0)}
	\sum_{\sa=\pm}w_{\lambda\sa}
	\left|\mathcal T_{\lambda\sa}\right|^2 \,. \label{eq: final_PowerSpectrum}
\end{align}
In Fig.~\ref{fig: tensorPS}, we show the time evolution of the tensor power spectrum for two helicities, constructed from the mode functions $\gamma_{\lambda\sa}(z)$, with the solid curves obtained by numerically solving the coupled equations \eqref{eq: EoM_Psi}. Since the initial conditions cannot be imposed at past infinity in the numerical calculation, we begin the evolution at $z_0=10^3$ and initialise the modes using the asymptotic BD behaviour in \eqref{eq: initial_condition}. The dashed lines show the exact final power spectrum given by the closed form expression \eqref{eq: final_PowerSpectrum}, in excellent agreement with the numerical results. The vertical black line marks horizon crossing at $z=1$, around which the two helicities begin to evolve differently. This behaviour follows from the parity odd contribution in \eqref{eq: EoM_Psi}, which is proportional to $z^{-1}$ and remains negligible in the far past before becoming important near horizon crossing.

\subsection{Spectator and Cross Power Spectrum}
The graviton power spectrum is the main observable of interest in this tensor sector, but the exact solution above contains the full information about both coupled fields. Indeed, the first equation in \eqref{eq: EoM_Psi} gives $\alpha Y_{\lambda\sa}=-\mathcal A X_{\lambda\sa}$, which together with the inverse relation $\sigma_{\lambda\sa}=-zY_{\lambda\sa}$, yields $\sigma_{\lambda\sa}=(z/\alpha)\mathcal A X_{\lambda\sa}$. We can therefore obtain the power spectrum of the spectator field and its cross spectrum with the graviton following a procedure similar to that used above.

Let us first consider the light mass range $0<m<3/2$, for which $\nu_t$ is real. Since we are interested in the late-time power spectrum, we need only the leading contribution in the superhorizon expansion. Applying $\mathcal A$ to the integral representation of $X_{\lambda\sa}$ then gives
\begin{align}
	\lim_{z\to 0}\sigma_{\lambda\sa}(z,k)
	&=\frac{H}{\sqrt{2k^3}}\sqrt{w_{\lambda\sa}}\,
	\mathcal S_{\lambda\sa}\,
	z^{\frac{3}{2}-\nu_t}\,,\label{eq:sigma-mode-late}
\end{align}
where
\begin{align}
	\mathcal{S}_{\lambda\sa}
	&=\frac{(-2i)^{\frac{3}{2}-\nu_t}e^{-\pi q_{\lambda\sa}/4}
	\Gamma(1-i\sa\rho)\Gamma(2\nu_t)}
	{\alpha\,\Gamma(\frac{-iq_{\lambda\sa}}{2})
	\Gamma\left(\frac{1}{2}+\nu_t-i\sa\rho\right)}\,.\label{eq:S-coefficient}
\end{align}
In contrast to the graviton mode, which approaches the constant at late times, the massive spectator mode decays outside the horizon as $z^{3/2-\nu_t}$. A detailed derivation of \eqref{eq:sigma-mode-late} is given in App.~\ref{app:sigma-cross-proof}.

Since the correlators are diagonal in helicity, we define the power spectrum of the spectator field and its cross spectrum with the graviton separately for each helicity. Following the dimensionless normalisation in \eqref{eq: h-powerS}, these are given by
\begin{align}
\begin{split}
	\left\langle\sigma_\lambda({\bf k})
	\sigma_\lambda({\bf k}')\right\rangle
	&=(2\pi)^3\,\delta^{(3)}({\bf k}+{\bf k}')\,\frac{\pi^2M_{\mathrm{pl}}^2}{2k^3}\,\mathcal P_\sigma^\lambda(z)\,,\\
	\left\langle h_\lambda({\bf k})\sigma_\lambda({\bf k}')\right\rangle
	&=(2\pi)^3\,\delta^{(3)}({\bf k}+{\bf k}')\,\frac{\pi^2M_{\mathrm{pl}}}{\sqrt{2}k^3}\,\mathcal P_{h\sigma}^\lambda(z)\,,
\end{split}
\end{align}
here the common time argument $z$ of the fields has been suppressed, and the factors of $M_{\mathrm{pl}}$ make both quantities dimensionless. Combining \eqref{eq:X-solution} with \eqref{eq:sigma-mode-late}, we find their leading late-time behaviour as
\begin{align}
\begin{split}
	\mathcal P_\sigma^\lambda(z)
	&={\mathcal P}_h^{(0)}z^{3-2\nu_t}
	\sum_{\sa=\pm}w_{\lambda\sa}
	\left|\mathcal S_{\lambda\sa}\right|^2\,,
	\\
	\mathcal P_{h\sigma}^\lambda(z)
	&=-{\mathcal P}_h^{(0)}z^{\frac{3}{2}-\nu_t}
	\sum_{\sa=\pm}w_{\lambda\sa}
	\mathcal T_{\lambda\sa}\mathcal S_{\lambda\sa}^{*}\,,
\end{split}
	\label{eq:sigma-cross-late}
\end{align}
where $\mathcal S_{\lambda\sa}$ and $\mathcal T_{\lambda\sa}$ are defined in \eqref{eq:S-coefficient} and \eqref{eq: 3F2expression}, respectively. The sum over $\sa$ in the spectator power spectrum can actually be performed explicitly, and we find
\begin{align}
	\mathcal P_\sigma^\lambda(z)
	&=\frac{2^{1-2\nu_t}\Gamma^2\left(2\nu_t\right)
	e^{-\lambda\pi\pv/2}}
	{\left|\Gamma\left(\frac{1}{2}+\nu_t-i\rho\right)\right|^2}\,z^{3-2\nu_t}{\mathcal P}_h^{(0)}\,.
	\label{eq:sigma-spectrum-final}
\end{align}
At equal times, the cross spectrum and the power spectrum of each field satisfy the Cauchy-Schwarz inequality
\begin{align}
	\left|\mathcal P_{h\sigma}^\lambda(z)\right|^2
	\leq\mathcal P_h^\lambda(z)\mathcal P_\sigma^\lambda(z)\,.
\end{align}

In the massless limit $m=0$, the spectator mode remains constant on superhorizon scales and continues to source the graviton at late times. Consequently, the graviton power spectrum and the cross spectrum diverge as $z\to0$. In the exact result, this behaviour appears through the pole of $\Gamma(\frac{3}{2}-\nu_t)$ in $\mathcal T_{\lambda\sa}$ at $\nu_t=3/2$, while $\mathcal S_{\lambda\sa}$ and the spectator power spectrum remain finite.

For $m>3/2$, $\nu_t$ becomes purely imaginary, so the two branches $z^{3/2-\nu_t}$ and $z^{3/2+\nu_t}$ both decay in magnitude as $z^{3/2}$ and must be included at leading order in \eqref{eq:sigma-mode-late}. Their interference produces oscillations in $\ln z$ in the spectator power spectrum with an envelope proportional to $z^3$, while the cross spectrum also oscillates in $\ln z$ with an envelope proportional to $z^{3/2}$.

\section{Special Cases and Limits}\label{sec:limits}
\noindent\textbullet\ \textit{Special mass case: $\Delta_t=0$.}\par
\vskip 8pt
\begin{figure}[htp]
\centering
\includegraphics[width=\columnwidth]{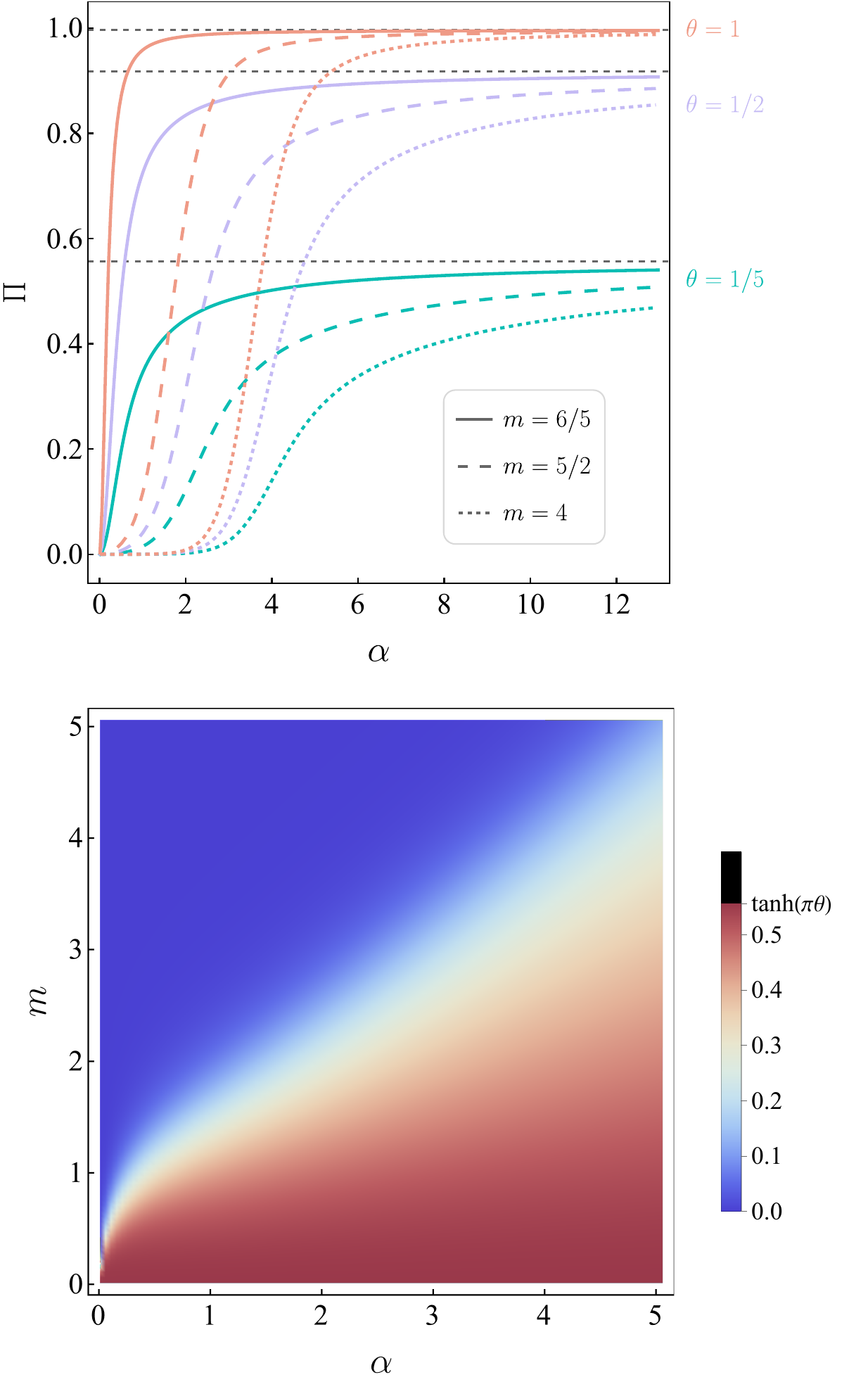}
\caption{\textit{Top}: Degree of circular polarisation \eqref{eq: chirality} as a function of the mixing strength $\alpha$, from the exact power spectrum \eqref{eq: final_PowerSpectrum}. Colours label the different parity-odd coupling $\pv$ and line styles the spin-2 mass. The gray dotted horizontal lines are the bound \eqref{eq: chirality_bound}. For each $\pv$ all masses approach the same bound once the mass term becomes negligible. \\
\textit{Bottom:} Scan over the parameter plane $(\alpha,m)$ at fixed $\pv=1/5$. All points satisfy the bound \eqref{eq: chirality_bound}, and the degree of circular polarisation \eqref{eq: chirality} decreases as the mass increases.
\label{fig: chirality}}
\end{figure}
As discussed above, when $\Delta_t=0$, or equivalently $\nu_t=1/2$, the coupled system can be diagonalised directly and the solution is greatly simplified. This special case therefore provides a useful check of our general result. In this limit, the first upper parameter of the hypergeometric function in \eqref{eq: 3F2expression} vanishes, reducing the function to unity. The power spectrum \eqref{eq: final_PowerSpectrum} then becomes
\begin{align}
	\mathcal{P}^{\lambda}_h={\mathcal P}_h^{(0)}\sum_{\sa=\pm}
	w_{\lambda\sa}\frac{1-e^{-\pi q_{\lambda\sa}}}{\pi{q}_{\lambda\sa}}\,.
\end{align}
	
Since $q_{\lambda-}<0$ and $q_{\lambda+}>0$ for $\alpha\neq0$, only the $\sa=-1$ channel is exponentially large, with $e^{-\pi q_{\lambda-}}=e^{\pi\rho}\,e^{-\pi\lambda\pv/2}$. For $\pv>0$ the negative helicity is therefore exponentially enhanced relative to the positive one, by $e^{2\pi\pv}$ in the exponent. To quantify the helicity asymmetry, we consider the degree of circular polarisation
\begin{align}
	\Pi\equiv\frac{\mathcal P^{-2}_h-\mathcal P^{+2}_h}
	{\mathcal P^{-2}_h+\mathcal P^{+2}_h} \,.\label{eq: chirality}
\end{align}
These two helicities satisfy the exact relation
\begin{align}
	\frac{e^{2\pi\pv}\mathcal P_h^{+2}-\mathcal P_h^{-2}}{{\mathcal P}_h^{(0)}}=\frac{4\pv e^{\pi\pv}}{\pi\alpha^2}\left[\cosh(\pi\rho)-\cosh(\pi\pv)\right] \,.
\end{align}
We have $\rho=\sqrt{\alpha^2+\pv^2}>\pv$, so the right-hand side is positive. It follows that $\mathcal P_h^{-2}/\mathcal P_h^{+2}<e^{2\pi\pv}$, which together with \eqref{eq: chirality} gives the upper bound
\begin{align}
	\Pi<\tanh(\pi\pv)\,.
	\label{eq: chirality_bound}
\end{align}
For $\alpha\gg\pv$, we have $\rho\simeq\alpha\gg\pv$, and the ratio approaches $e^{2\pi\pv}$, so that $\Pi\to\tanh(\pi\pv)$. 
	
This bound \eqref{eq: chirality_bound} was derived for $\Delta_t=0$, it is  saturated in the massless limit $m=0$, for which $\Pi=\tanh(\pi\pv)$.\footnote{Strictly speaking, the power spectrum obtained from \eqref{eq: 3F2expression} diverges in the massless limit because of the factor $\Gamma({3}/{2}-\nu_t)$. Since the divergent factor is common to both helicities, it cancels when forming the ratio in \eqref{eq: chirality}, allowing us to define it formally and discuss the bound in this limit.} For a general mass, we expect the same bound to hold, but proving this rigorously requires deriving inequalities for the higher order hypergeometric functions, which is a mathematical problem beyond the scope of this work. The physical picture can still be understood directly from the equation of motion \eqref{eq: EoM_Psi}. The exponential enhancement is generated around $z\simeq|q_{\lambda-}|\sim\rho$, where the effective frequency squared $1+q_{\lambda-}/z$ of the amplified channel vanishes. At this scale, the ratio of the mass term to the mixing term is $|\Delta_t|/\rho^2$, so once $\rho^2\gg|\Delta_t|$, the mass term becomes negligible and the dynamics reduces to the special mass case discussed above, recovering the bound \eqref{eq: chirality_bound}. In the opposite regime, a sufficiently large positive $\Delta_t$ shortens the growth band of the negative helicity modes, suppressing their enhancement and reducing the difference between the two helicities.

The origin of this bound can also be understood directly from the spectator field. We define its degree of circular polarisation $\Pi_\sigma$ in the same way as \eqref{eq: chirality}, replacing $\mathcal P_h^\lambda$ with $\mathcal P_\sigma^\lambda$. Using the spectator power spectrum \eqref{eq:sigma-spectrum-final}, we find
\begin{align}
	\frac{\mathcal P_\sigma^{-2}(z)}{\mathcal P_\sigma^{+2}(z)}
	=e^{2\pi\pv}\,,
\end{align}
which gives $\Pi_\sigma=\tanh(\pi\pv)$. Remarkably, this is exactly the upper bound on $\Pi$ in \eqref{eq: chirality_bound}. Since the parity asymmetry of the graviton is generated through its mixing with the spectator field, its chirality is naturally bounded by that of the spectator, giving $\Pi<\Pi_\sigma$.

To test the bound for general masses, in the top panel of Fig.~\ref{fig: chirality} we plot $\Pi$ as a function of $\alpha$ for several values of $\pv$, with the line styles distinguishing different masses. For each $\pv$, the curves remain below the limiting value $\tanh(\pi\pv)$, showing clearly that the bound is determined only by $\pv$ and does not depend on either the mixing strength or the mass. In the bottom panel, we fix $\pv=1/5$ and scan the $(m,\alpha)$ plane. As shown in the colour bar, to make any violation immediately visible, we intentionally assign the colour black to values above the bound, but no such points are found throughout the parameter range. Moreover, at fixed $\alpha$, $\Pi$ increases as the mass is lowered and approaches the massless result, which exactly saturates the bound. 
	
These results provide evidence that the bound \eqref{eq: chirality_bound} holds for general masses, and we stress that it is intrinsic to the general tensor system considered here and is expected to be satisfied throughout its all parameter space. In a complete model, the allowed parameters may be further restricted by other considerations, including the stability condition for the helicity-0 component discussed earlier and constraints from backreaction \cite{Gorji:2023cmz}, but we do not include them in the present analysis since they depend on the specific model under consideration.

\vskip 8pt
\noindent\textbullet\ \textit{Parity symmetric case: $\pv=0$.}\par
\vskip 8pt
When the parity odd term is switched off, $\pv=0$, the two helicities become identical, with $q_{\lambda\sa}=\sa\alpha=\sa\rho$. In this limit, the hypergeometric function ${}_3F_2$ reduces to a ratio of Gamma functions, and the power spectrum becomes
\begin{align}
	\mathcal{P}^{\lambda}_{h}={\mathcal P}_h^{(0)} \left|\frac{\Gamma\left(\frac{3}{4}-\frac{\nu_t}{2}\right)\Gamma\left(\frac{3}{4}+\frac{\nu_t}{2}\right)}{\Gamma\left(\frac{3}{4}+\frac{\nu_t}{2}+\frac{i\alpha}{2}\right)\Gamma\left(\frac{3}{4}-\frac{\nu_t}{2}+\frac{i\alpha}{2}\right)}\right|^2 \,, \label{eq:PS_parityeven}
	\end{align}
which is in complete agreement with the corresponding result for the scalar case obtained in \cite{Huenupi:2026abj}. For the special mass $\Delta_t=0$, or equivalently $\nu_t=1/2$, this expression further reduces to $\mathcal{P}^{\lambda}_{h}={\mathcal P}_h^{(0)}\sinh(\pi\alpha)/(\pi\alpha)$. Finally, taking the limit $\alpha\to0$ in \eqref{eq:PS_parityeven} recovers the free graviton spectrum per helicity $\mathcal{P}^{\lambda}_{h}={\mathcal P}_h^{(0)}$ which is defined in \eqref{eq:PS-free}.
	
\vskip 8pt
\noindent\textbullet\ \textit{Weak mixing limit: $\mathcal{O}(\alpha^2)$.}\par
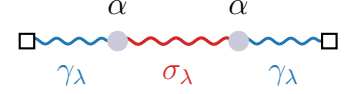
\begin{figure}[h!]
		\centering
		\begin{tikzpicture}[x=1cm,y=1cm]
			\tikzset{
				graviton/.style={
					draw=blue3,
					line width=1.1pt,
					decorate,
					decoration={snake,amplitude=1.1pt,segment length=8.5pt}
				},
				spectator/.style={
					draw=red3,
					line width=1.2pt,
					decorate,
					decoration={snake,amplitude=1.2pt,segment length=9.5pt}
				}
			}
			\coordinate (leftboundary) at (-2,0);
			\coordinate (leftvertex) at (-0.8,0);
			\coordinate (rightvertex) at (0.8,0);
			\coordinate (rightboundary) at (2,0);
			
			\draw[graviton] (leftboundary) -- (leftvertex);
			\draw[spectator] (leftvertex) -- (rightvertex);
			\draw[graviton] (rightvertex) -- (rightboundary);
			\draw[draw=lightgray2,fill=lightgray2]
			(leftvertex) circle (0.1296cm);
			\draw[draw=lightgray2,fill=lightgray2]
			(rightvertex) circle (0.1296cm);
			\draw[line width=0.8pt,fill=white]
			(-2.096,-0.096) rectangle (-1.904,0.096);
			\draw[line width=0.8pt,fill=white]
			(1.904,-0.096) rectangle (2.096,0.096);
			
			\node[above=7pt,font=\large] at (leftvertex) {$\alpha$};
			\node[above=7pt,font=\large] at (rightvertex) {$\alpha$};
			\node[below=6pt,text=blue3,font=\large] at (-1.4,0) {$\gamma_\lambda$};
			\node[below=6pt,text=red3,font=\large] at (0,0) {$\sigma_\lambda$};	
			\node[below=6pt,text=blue3,font=\large] at (1.4,0) {$\gamma_\lambda$};
		\end{tikzpicture}
		\caption{Leading single-exchange contribution to the graviton power spectrum at $\mathcal{O}({\alpha}^2)$. The blue wavy lines denote graviton propagators, while the red wavy line denotes the propagator of the helicity-2 component of the spin-2 field $\sigma_{ij}$.\label{fig: SK_exchange}}
	\end{figure}

\vskip 8pt
When the quadratic mixing is sufficiently weak, the graviton correlators can be computed perturbatively. The leading correction to the power spectrum comes at $\mathcal{O}(\alpha^2)$ from the single-exchange process shown in Fig.~\ref{fig: SK_exchange}.
Such a diagram can be evaluated perturbatively using the standard in-in formalism or the SK formalism. Since two interactions already give a factor of $\alpha^2$, we can simply ignore the $\alpha$ dependence of the mass term in \eqref{eq: sigma_action} when computing the result at $\mathcal{O}(\alpha^2)$, as it contributes only at higher orders. The mode function of the $\sigma_{ij}$ field with helicity $\lambda$ is then given by \cite{Stefanyszyn:2023qov}
\begin{align}
	\sigma_\lambda(\tau,k)
	&=-\frac{H\tau}{\sqrt{2k}}\,
	e^{-\frac{\pi\chi_\lambda}{2}}W_{i\chi_\lambda,\,\nu_t}(2ik\tau)\,,
\end{align}
where 
\begin{align}
	\chi_\lambda\equiv\frac{\lambda\pv}{2}\,,\qquad \chi_{\pm2}=\pm\pv\,.
\end{align}
The SK propagators represented by the red wavy line in Fig.~\ref{fig: SK_exchange} can then be constructed from this mode function as \cite{Chen:2017ryl}
\begin{align}
\begin{split}
	D^\lambda_{-+}(k;\tau_1,\tau_2)&=\sigma_\lambda(\tau_1,k)\sigma_\lambda^*(\tau_2,k)\,,
	\\
	D^\lambda_{+-}(k;\tau_1,\tau_2)&=\sigma^*_\lambda(\tau_1,k)\sigma_\lambda(\tau_2,k)\,,
	\\
	D^\lambda_{++}(k;\tau_1,\tau_2)
	&=\Theta(\tau_1-\tau_2)D^\lambda_{-+}
	+\Theta(\tau_2-\tau_1)D^\lambda_{+-}\,,
	\\
	D^\lambda_{--}(k;\tau_1,\tau_2)
	&=\Theta(\tau_1-\tau_2)D^\lambda_{+-}+\Theta(\tau_2-\tau_1)D^\lambda_{-+}\,,
\end{split}
\end{align}
here the labels $+$ and $-$ denote forward and backward branches of the SK path, respectively. The blue wavy lines ending on the late-time boundary in Fig.~\ref{fig: SK_exchange} correspond to the graviton bulk-to-boundary propagators, which are
\begin{align}
	G_{\sa}(k,\tau)&=\frac{H^2}{2k^3}
	(1-i\sa k\tau)e^{i\sa k\tau}\,,
	\qquad \sa=\pm1\,.
\end{align}
Combining these ingredients with the interaction \eqref{eq:int}, the leading contribution from the exchange of $\sigma_{ij}$ can be written as
\begin{align}
	c_{\lambda,\mathrm{SK}}^{(2)}
	\equiv
	\frac{\left.\delta\mathcal P_h^\lambda\right|_{\mathrm{SK}}}{{\mathcal P}_h^{(0)}\alpha^2}
	=\frac{1}{2}\sum_{\sa,\sb=\pm1}
	\mathcal I^\lambda_{\sa\sb}\,,
	\label{eq: SK_power_correction}
\end{align}
where all time integrals have been collected into the dimensionless quantity
\begin{align}
	\mathcal{I}^{\lambda}_{\sa\sb}=-
	\frac{\sa\sb k}{H^2}
	\int_{-\infty}^{0}\frac{\dd\tau_1}{\tau_1^2}\frac{\dd\tau_2}{\tau_2^2}
	e^{i(\sa \tau_1+\sb\tau_2)k}\,
	D^{\lambda}_{\sa\sb}(k;\tau_1,\tau_2)\,,
\end{align}
which is precisely the Whittaker seed integral evaluated in \cite{Qin:2023closed}. After translating their result to our conventions and analytically continuing in $\nu_t$, the non-time-ordered contribution becomes
	\begin{align}
		\mathcal I^{\lambda}_{\sa,-\sa}
		=\frac{e^{-\pi\chi_\lambda}}{2}
		\frac{\Gamma^2\left(\frac{1}{2}-\nu_t\right)\Gamma^2\left(\frac{1}{2}+\nu_t\right)}{\Gamma\left(1+i\sa\chi_\lambda\right)
			\Gamma\left(1-i\sa\chi_\lambda\right)}\,,
		\label{eq: SK_opposite_branch}
	\end{align}
whereas the (anti-)time-ordered contribution is\footnote{Our integral is obtained from the final expression of \cite{Qin:2023closed} by setting $p_1=p_2=-1$. We have also corrected a typo in the sign of one of the $\Gamma$ function arguments.}
\begin{align}
\begin{split}
		\mathcal I^{\lambda}_{\sa\sa}
		&=\frac{i\sa\pi}{4}
		\frac{e^{-2\pi\chi_\lambda}+e^{2\pi i\nu_t}}{\cos^2(\pi\nu_t)\Gamma^2\left(1+i\sa\chi_\lambda\right)}
		\\
		&\times
		\Gamma\left(\frac{1}{2}-\nu_t+i\sa\chi_\lambda\right)\Gamma\left(\frac{1}{2}+\nu_t+i\sa\chi_\lambda\right)
		\\
		&-\frac{2}{\left(1+2\sa\nu_t\right)^2}\,\pFq{3}{2}{1,\frac{1}{2}+\sa\nu_t-i\sa\chi_\lambda,1}
		{\frac{3}{2}+\sa\nu_t,\frac{3}{2}+\sa\nu_t}
		{1}\,.
\end{split}
		\label{eq: SK_same_branch}
	\end{align}
Summing over $\sa=\pm1$ in \eqref{eq: SK_opposite_branch} and \eqref{eq: SK_same_branch} gives all four SK contributions. Although the individual same branch integrals are complex, their total contribution to the power spectrum is real.
	
We can now compare this result with our exact expression for the power spectrum in \eqref{eq: final_PowerSpectrum}. Since it is invariant under $\alpha\to-\alpha$, its weak coupling expansion contains only even powers of $\alpha$ and can be written as
\begin{align}
    \sum_{\sa=\pm1}w_{\lambda\sa}
    \left|\mathcal T_{\lambda\sa}\right|^2
    =1+\alpha^2c_{\lambda,\mathrm{exact}}^{(2)}
    +\mathcal O(\alpha^4)\,.
\end{align}
Here the expansion is performed only after summing over the two eigenchannels. Evaluating two analytical expressions numerically, we find
\begin{align}
	c_{\lambda,\mathrm{exact}}^{(2)}=c_{\lambda,\mathrm{SK}}^{(2)}\,.
	\label{eq: exact_SK_comparison2}
\end{align}
This agreement confirms that the weak coupling expansion of the exact result reproduces the SK result \eqref{eq: SK_power_correction} at  $\mathcal{O}(\alpha^2)$. An analytical proof would require some additional steps, since expanding the exact result generates derivatives of the hypergeometric functions with respect to their parameters. These derivatives can be reduced using identities among different hypergeometric functions. As our aim here is to perform a consistency check, we instead compare the two analytical expressions in \eqref{eq: exact_SK_comparison2} numerically for several representative parameter choices.\footnote{For this comparison, we work in the heavy mass cases to avoid possible IR divergences in the perturbative calculation.}

The complete expressions for the graviton power spectrum at $\mathcal{O}(\alpha^2)$ involve generalised hypergeometric functions, but they simplify when we take the difference between two helicities, which precisely captures the parity violation effect. To see this, we first collect the hypergeometric contribution in \eqref{eq: SK_same_branch} into the function
\begin{align}
	\mathcal{F}_{\lambda}
	\equiv\sum_{\sa=\pm1}\frac{2}{\left(1+2\sa\nu_t\right)^2}\pFq{3}{2}{1,\frac{1}{2}+\sa\nu_t-i\sa\chi_\lambda,1}
	{\frac{3}{2}+\sa\nu_t,\frac{3}{2}+\sa\nu_t}{1}\,.
\end{align}
One can verify numerically that the difference between the two helicities satisfies	
\begin{align}
	\frac{\mathcal{F}_{-2}-\mathcal{F}_{+2}}{2\pi i\tan(\pi\nu_t)}=\text{Im}\left[\frac{\Gamma\left(\frac{1}{2}-\nu_t+i\pv\right)\Gamma\left(\frac{1}{2}+\nu_t+i\pv\right)}{\Gamma^2\left(1+i\pv\right)}\right]\,,
\end{align}
this simplification follows from the general theorems established in \cite{Stefanyszyn:2023qov,Stefanyszyn:2024msm,Stefanyszyn:2025yhq}, which show that parity odd correlators have a much simpler structure. Combining this identity with \eqref{eq: SK_opposite_branch} and \eqref{eq: SK_same_branch}, the corresponding helicity difference takes the form
\begin{align}
	c^{(2)}_{-2}-c^{(2)}_{+2}
	=\frac{\pi}{\mathcal{R}^2\pv}\left[
	1-\sqrt{1+\mathcal{R}^2}\sin\varphi\right]\,, \label{eq: helicity_identity}
\end{align}
where the ratio $\mathcal{R}$ and the phase $\varphi$ are defined as	
\begin{align}
	\mathcal{R}(\pv,\nu_t)&\equiv \frac{\cos(\pi\nu_t)}{\sinh(\pi\pv)}\,,\nonumber\\
	\varphi(\pv,\nu_t)&\equiv\arg\left[\frac{\Gamma^2\left(1+i\pv\right)}{\Gamma\left(\frac{1}{2}-\nu_t+i\pv\right)
	\Gamma\left(\frac{1}{2}+\nu_t+i\pv\right)}\right]\,.
\end{align}
It would be interesting to understand how the perturbative helicity difference \eqref{eq: helicity_identity} is related to the bound in \eqref{eq: chirality_bound}, which holds for arbitrary $\alpha$, and we leave this question for future work.

\section{Summary}\label{sec:summary}
We have studied the coupled tensor perturbations of the metric and a spectator spin-2 field during inflation. The two fields mix linearly through the operator $\sigma_{ij}\delta K_{ij}$, while the spectator sector contains a parity-odd effect parametrised by $\pv$. Instead of treating the quadratic mixing perturbatively, we have solved the coupled system exactly for arbitrary mass and mixing strength $\alpha$. At the special mass $\Delta_t=0$, the two eigenchannels of the mixing matrix decouple and their mode functions are given by Whittaker functions. For a general mass, no constant field rotation diagonalises the system. Nevertheless, the exact solutions can be constructed by acting on the reference Whittaker modes with the operator-valued Gauss hypergeometric function $\mathcal D_\sa(\hat{x})$ defined in \eqref{eq: D-def}. Since this operator reduces to the identity in the far past, the construction preserves the BD normalisation.

Using these exact mode functions, we have obtained the late-time power spectrum of each graviton helicity in closed form, as given in \eqref{eq: final_PowerSpectrum}, in terms of the generalised hypergeometric function ${}_3F_2$. The parity-odd term makes the eigenvalues $q_{\lambda\sa}$ helicity dependent and thereby generates a chiral spectrum. For $\pv>0$, the negative helicity is exponentially enhanced relative to the positive one, while the total amplitude grows rapidly with the mixing strength $\alpha$. Large mixing can therefore produce a large and highly chiral primordial GW spectrum. For $\Delta_t=0$, we have explicitly derived the bound \eqref{eq: chirality_bound} on the degree of circular polarisation defined in \eqref{eq: chirality}, and our numerical results provide evidence that it also holds for a general mass. The bound is formally saturated in the massless limit, where $\Pi=\tanh(\pi\pv)$ independently of $\alpha$, and is approached for any mass when the mixing becomes sufficiently large.

Using the spectator mode functions, we have also obtained the leading late-time power spectrum of the spectator field and its cross spectrum with the graviton. For $0<m<3/2$, the spectator degree of circular polarisation is $\Pi_\sigma=\tanh(\pi\pv)$, which is precisely the upper limit on the graviton polarisation. This relation gives a physical interpretation of the bound in terms of the helicity asymmetry of the spectator field that sources the graviton through the mixing.

We have tested the exact result both numerically and in several limiting cases. The closed-form spectrum agrees with direct numerical integration of the coupled equations. In the parity-symmetric limit, it reproduces the exact scalar result of \cite{Huenupi:2026abj}, while in the weak-mixing limit it agrees with the perturbative single-exchange SK computation at $\mathcal O(\alpha^2)$. These checks confirm that the solution consistently interpolates between the perturbative and strongly mixed regimes.

Since the quadratic mixing is resummed to all orders in $\alpha$, the exact mode functions derived here also provide the building blocks for computing higher point correlation functions when a perturbative expansion in $\alpha$ is no longer reliable. A natural extension is therefore to study graviton non-Gaussianities within the same nonperturbative treatment of linear mixing, in analogy with recent studies of the scalar bispectrum at strong mixing \cite{Huenupi:2026aqc,Wang:2026lff,Pinol:2026xnl,Belrhali:2026uxn}. Another important step is to evolve the enhanced chiral primordial spectrum to late times and investigate its possible signatures in GW observations at frequencies above CMB scales.

\vskip 5pt
\paragraph*{Acknowledgements}
This work is supported by IBS under the project code IBS-R018-D3. OpenAI ChatGPT and Anthropic Claude were used to assist with language editing and the discussion of some analytical and numerical calculations. All results were independently verified by the authors, who take full responsibility for the scientific content of the manuscript.
	
\vskip 5pt

\appendix
\section{Spectator and Cross Power Spectrum}
\label{app:sigma-cross-proof}

In this appendix, we derive the expressions for the late-time spectator mode and cross spectrum presented in the main text. We first restrict to the light mass range $0<m<3/2$, for which $\nu_t$ is real and positive, and discuss the heavy mass range at the end.

To derive the late-time behaviour of the spectator mode, we begin with $\sigma_{\lambda\sa}=(z/\alpha)\mathcal{A} X_{\lambda\sa}$. Using \eqref{eq: general_sol} together with the integral representation \eqref{eq:u-int-rep}, we again express $X_{\lambda\sa}$ in a form in which all the $z$-dependence is carried by $e^{iz(1+2t)}$. Acting with $\mathcal A$ then gives
\begin{align}
	\mathcal{A}\,e^{iz(1+2t)}
	&=-4z\,t(1+t)\,e^{iz(1+2t)}\,,
\end{align}
the resulting factor  cancels $t^{-1}(1+t)^{-1}$ in the integrand, giving
\begin{align}
\begin{split}
	\sigma_{\lambda\sa}(z,k)
	=&-\frac{4Hz^2}{\alpha\sqrt{2k^3}}\,
	\frac{\sqrt{w_{\lambda\sa}\,}e^{iz-\pi q_{\lambda\sa}/4}}{\Gamma(-iq_{\lambda\sa}/2)}\\
	&\times\int_0^\infty\dd t\,\left(1+\frac{1}{t}\right)^{iq_{\lambda\sa}/2}
	\mathcal{D}_\sa(-t)\,e^{2izt}\,.
\end{split}
\end{align}
To evaluate the integral as a convergent series, we first use a transformation of the hypergeometric function \cite{NIST:DLMF} to write
\begin{align}
\begin{split}
	&\mathcal D_\sa(-t)\\
	&=(1+t)^{\nu_t-\frac{1}{2}}\pFq{2}{1}{\frac{1}{2}-\nu_t,\frac{1}{2}-\nu_t-i\sa\rho}{1-i\sa\rho}{\frac{t}{1+t}}\,.
\end{split}
\end{align}
Since $0\leq t/(1+t)<1$ throughout the integration domain, we can expand the transformed hypergeometric function inside the integral as
\begin{align}
	&\mathcal D_\sa(-t)\nonumber\\
	&=(1+t)^{\nu_t-\frac{1}{2}}\sum_{n=0}^\infty\frac{(\frac{1}{2}-\nu_t)_n(\frac{1}{2}-\nu_t-i\sa\rho)_n}{(1-i\sa\rho)_n\,n!}\left(\frac{t}{1+t}\right)^n\,.
\end{align}
Here $(a)_n\equiv\Gamma(a+n)/\Gamma(a)$ denotes the Pochhammer symbol. The difference between the lower parameter and the sum of the two upper parameters is $2\nu_t>0$, so the series remains absolutely convergent as $t\to\infty$. With the usual $i\epsilon$ prescription, each term can then be integrated in terms of the confluent hypergeometric function of the second kind $U$, and we obtain
\begin{align}
\begin{split}
	&\sigma_{\lambda\sa}(z,k)
	=-\frac{4Hz^2}{\alpha\sqrt{2k^3}}\,
	\frac{\sqrt{w_{\lambda\sa}\,}e^{iz-\pi q_{\lambda\sa}/4}}
	{\Gamma(-iq_{\lambda\sa}/2)}\\
	&\times\sum_{n=0}^\infty\frac{(\frac{1}{2}-\nu_t)_n(\frac{1}{2}-\nu_t-i\sa\rho)_n}{(1-i\sa\rho)_n\,n!}\,\,\mathcal{U}_{\lambda\sa,n}(z)\,,
\end{split}
\end{align}
where we have introduced the shorthand notation
\begin{align}
\begin{split}
	\mathcal{U}_{\lambda\sa,n}(z)
	&\equiv\Gamma\left(n+1-\frac{iq_{\lambda\sa}}{2}\right)
	\\
	&\times U\left(n+1-\frac{iq_{\lambda\sa}}{2},\frac{3}{2}+\nu_t,-2iz\right)\,.
\end{split}
\end{align}
With the full series solution, we can directly extract its leading late-time behaviour from the small-$z$ expansion of $\mathcal{U}_{\lambda\sa,n}(z)$
\begin{align}
	\mathcal U_{\lambda\sa,n}(z)
	&\sim
	\Gamma\left(\frac{1}{2}+\nu_t\right)
	(-2iz)^{-\frac{1}{2}-\nu_t}\,.
\end{align}
Since the leading term is independent of $n$, the remaining series can be summed as
\begin{align}
\begin{split}
	&\sum_{n=0}^\infty
	\frac{(\frac{1}{2}-\nu_t)_n
	(\frac{1}{2}-\nu_t-i\sa\rho)_n}
	{(1-i\sa\rho)_n\,n!}
	\\
	&\qquad=\frac{\Gamma(1-i\sa\rho)\Gamma(2\nu_t)}
	{\Gamma\left(\frac{1}{2}+\nu_t\right)
	\Gamma\left(\frac{1}{2}+\nu_t-i\sa\rho\right)}\,.
\end{split}
\end{align}
Combining all the expressions above, we now find the leading late-time spectator mode as
\begin{align}
\begin{split}
	&\lim_{z\to 0}\sigma_{\lambda\sa}(z,k)
	=\frac{H}{\sqrt{2k^3}}\sqrt{w_{\lambda\sa}}\,
	z^{\frac{3}{2}-\nu_t}\\
	&~~~~~~\times\frac{(-2i)^{\frac{3}{2}-\nu_t}e^{-\pi q_{\lambda\sa}/4}\Gamma(1-i\sa\rho)\Gamma(2\nu_t)}
	{\alpha\,\Gamma(\frac{-iq_{\lambda\sa}}{2})
	\Gamma\left(\frac{1}{2}+\nu_t-i\sa\rho\right)}\,.
\end{split}
\end{align}
This is the late-time result quoted in the main text \eqref{eq:sigma-mode-late}. Together with the late-time graviton mode following from \eqref{eq: inverse_relation} and \eqref{eq:X-solution}, this result gives the spectator power spectrum \eqref{eq:sigma-spectrum-final} and the cross spectrum \eqref{eq:sigma-cross-late}.

We finally turn to the heavy mass range $m>3/2$, for which $\nu_t$ is purely imaginary. In this case, the series above is no longer absolutely convergent at unity, so the small-$z$ limit cannot be interchanged with the sum. This is why the derivation above gives only the $z^{3/2-\nu_t}$ contribution. Since the exact mode is symmetric under $\nu_t\to-\nu_t$, the $z^{3/2+\nu_t}$ contribution is obtained by sending $\nu_t\to-\nu_t$ in the result above. The two contributions are of the same order and must both be included at leading order.

\bibliography{Refs}
\bibliographystyle{utphys}

\end{document}